\documentclass[conference,a4paper]{isap2021}
\usepackage{graphicx}
\usepackage{color}
\usepackage{multirow}

\title{
Enhancing and Localizing Surface\\Wave Propagation with Reconfigurable Surfaces
}

\author{
Zhiyuan Chu$\stackrel{\S}{,}$ Kai-Kit Wong, and Kin-Fai Tong
\\
Department of Electronic and Electrical Engineering, University College London
\\
Torrington Place, London, United Kingdom
\\
$^\S$Corresponding author, email: $\rm zhiyuan.chu.18@ucl.ac.uk$
}

\begin{document}

\baselineskip 4.5mm

\maketitle

\begin{abstract}
As an attempt to develop a reconfigurable surface architecture that can use liquid metal such as Galinstan to shape surface channels on demand, this paper considers a punctured surface where cavities are evenly distributed and can be filled with liquid metal potentially via digitally controlled pumps. In this paper, we look at the benefits of such architecture in terms of surface-wave signal enhancement and isolation, and examine how various system parameters impact the performance using full wave 3-dimensional electromagnetic simulations. It is shown that extraordinary signal shaping can be obtained.
\end{abstract}



\section{Introduction}
Intelligent surfaces are anticipated by many to provide the next leap in 6G mobile communications \cite{Akyildiz-18,Marco-19}. While most focused on utilizing surfaces as intelligent passive reflectors, the authors in \cite{Wong-21} presented a vision of benefiting from the desirable propagation characteristics of surface waves \cite{sarkar2017surface} and creating a smart radio environment. The vision is attractive but only if adaptable surface waves are realizable. $y_r$

Of particular relevance is the work in \cite{gao2018surface} where mechanically controlled metal bars were made to appear or disappear to create changeable surface wave pathways. The practicality of the approach in \cite{gao2018surface} is nonetheless questionable. To address this, \cite{Chu-aps21} proposed to use liquid metal bars to form specific pathways to shape surface waves. Unfortunately, the setup in \cite{Chu-aps21} did not consider the materials for containing the liquid metal or they were assumed to be infinitesimally thin, which was unrealistic or practically unachieveable.

\begin{figure}[]
\centering
\includegraphics[width=7.5cm]{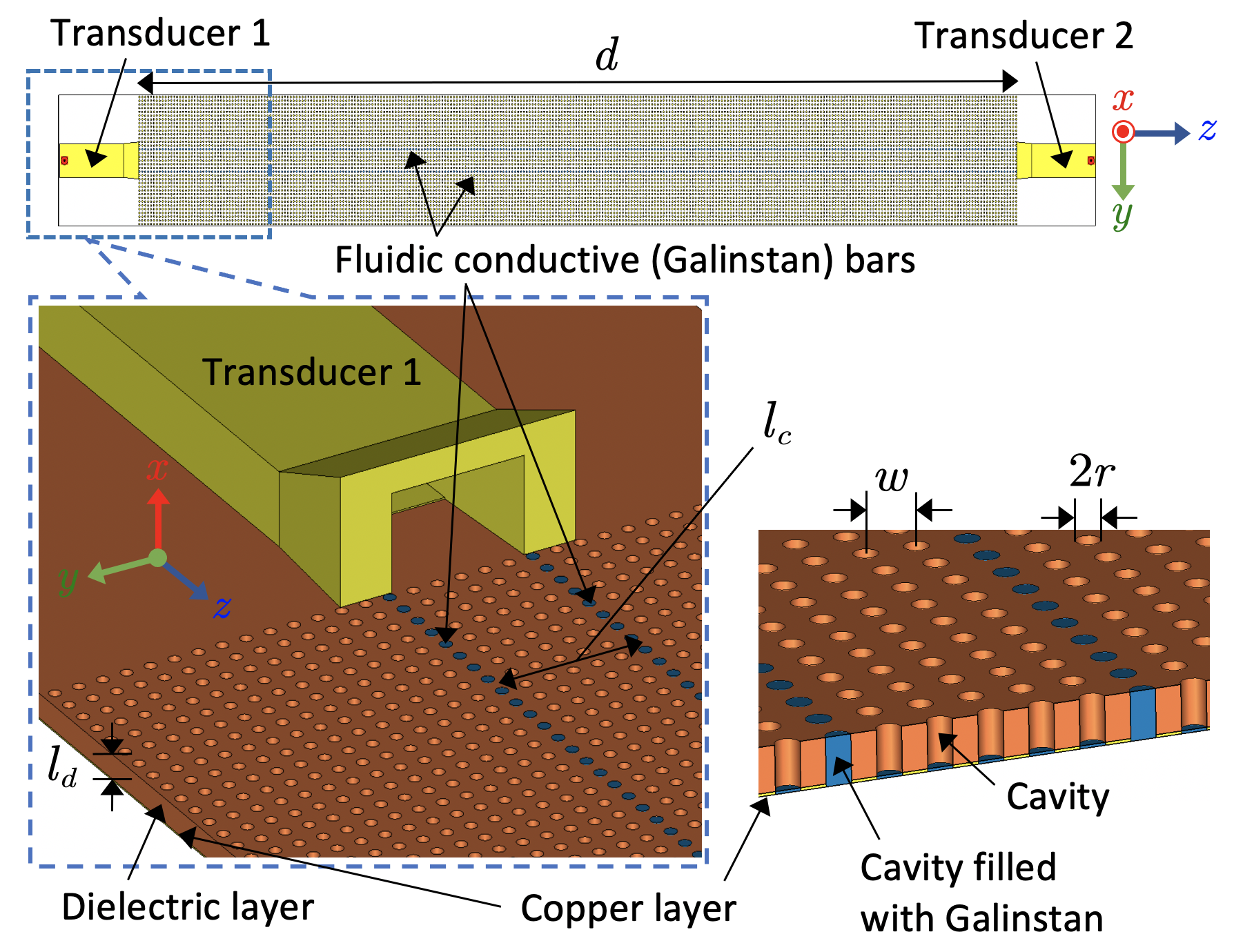}
\caption{The punctured surface model with cavities potentially filled with liquid metal, e.g., Galinstan, on demand via digitally controlled pumps.}\label{fig:Model}
\end{figure}

In this paper, we propose a punctured surface model, as shown in Fig.~\ref{fig:Model}, where cavities are evenly distributed and can be filled with liquid metal on demand to form specific pathways for surface wave propagation. Surface wave signal localization and enhancement will be investigated.

\section{The Punctured Surface Model}
It was illustrated in \cite{Chu-aps21} that liquid metal structures can be used to form isolated pathways for surface wave propagation but their platform does not readily facilitate implementation since it would be difficult to make the tiny tubes with sub-millimeter thickness. The proposed model in this paper thus opts for a punctured surface where cavities are formed to be filled with liquid metal such as Galinstan to reproduce the ability to shape the transmitted surface waves. 

Specifically, we have the parameters of the model defined in Fig.~\ref{fig:Model} and their numerical values presented in TABLE \ref{tab:setup}. With a porosity $\phi$, the effective relative dielectric constant of the surface can be found as \cite{Liu-16}
\begin{equation}
\varepsilon_r^{\rm eff}=\frac{\varepsilon_r\left[1+3\varepsilon_r+3\phi(1-\varepsilon_r)\right]}{1+3\varepsilon_r +\phi(\varepsilon_r-1)},
\end{equation}
where $\varepsilon_r$ denotes the original relative dielectric constant. In what follows, the surface impedance can be obtained by
\begin{equation}
X_s=2\pi f\mu_0\left(\frac{\varepsilon_r^{\rm eff}-1}{\varepsilon_r^{\rm eff}}l_d+\frac{\Delta}{2}\right),
\end{equation}
where $\Delta$ is the skin depth of the metal sheet, $f$ is the operating frequency, and $\mu_0$ is the permeability of the free space. In our model, we have $4$ cavities for every $4{\rm mm}\times 4{\rm mm}$ space, which gives $\phi=\frac{4\pi (0.5)^2}{4^2}=19.63\%$. Also, a fixed transducer with height of $2.8{\rm mm}$ and width of $9.6{\rm mm}$ is used as this paper's focus is on the design of the reconfigurable surface. That said, as in \cite{Tong-19}, it is possible to choose the dimensions of the transducer more carefully to have the optimal surface impedance with maximum excitation efficiency.

\begin{table}[]
\begin{center}
\begin{tabular}{r||c}
{\bf Parameter} & {\bf Value}\\
\hline
\mbox{radius of the bar}, $r$ & $0.5{\rm mm}$\\
\hline
\mbox{center-to-center separation between bars}, $w$ & $2{\rm mm}$\\
\hline
\mbox{thickness of the dielectric layer}, $l_d$ & $1.6{\rm mm}$\\

\hline
\mbox{channel propagation distance}, $d$ & $600{\rm mm}$\\
\hline
\mbox{channel width}, $l_c$ & $12, 16, 20{\rm mm}$\\
\hline
\mbox{operating frequency}, $f$ & $30, 40{\rm GHz}$\\
\hline
\mbox{inductive surface impedance}, $X_s$ & $j184\Omega$ at $30{\rm GHz}$\\
\mbox{} & $j245\Omega$ at $40{\rm GHz}$\\
\hline
\mbox{dielectric porosity}, $\phi$ & $19.63\%$\\
\hline
\mbox{dielectric effective relative permittivity}, $\varepsilon _r^{\rm eff}$ & $1.94$\\
\hline
\mbox{conductivity for Galinstan}, $\sigma_{\rm g}$ & $3.46\times 10^6~{\rm Sm}^{-1}$\\
\hline
\mbox{conductivity for copper}, $\sigma_{\rm c}$ & $59.6\times 10^6~{\rm Sm}^{-1}$
\end{tabular}
\end{center}
\caption{The parameters used in the simulations.}\label{tab:setup}
\vspace{-5mm}
\end{table}


\begin{figure*}[]
\centering
\includegraphics[width=15cm]{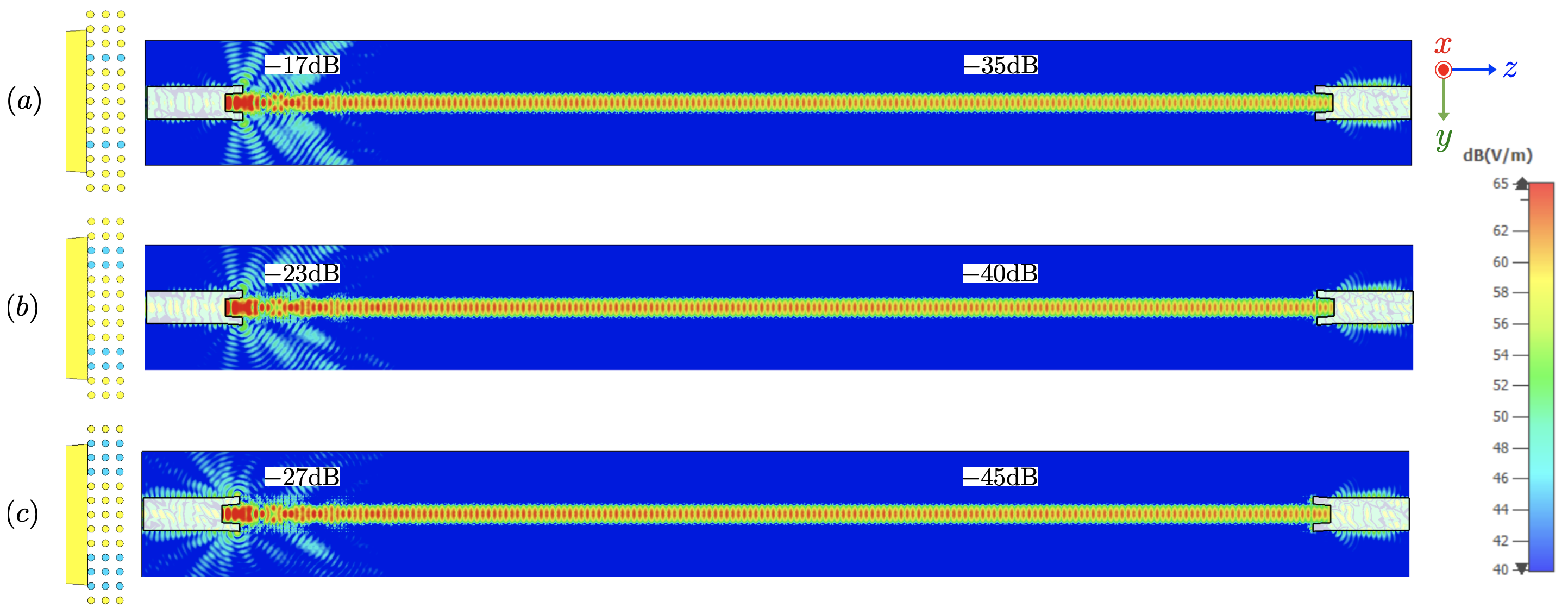}
\caption{The E-field distribution in dB over the surface with $l_c=12{\rm mm}$ and $f=30{\rm GHz}$ when shaping the path with $(a)$ one layer, $(b)$ two layers or $(c)$ three layers of cavities filled with Galinstan.}\label{fig:layer}
\end{figure*}

\begin{figure}[]
\centering
\includegraphics[width=8.5cm]{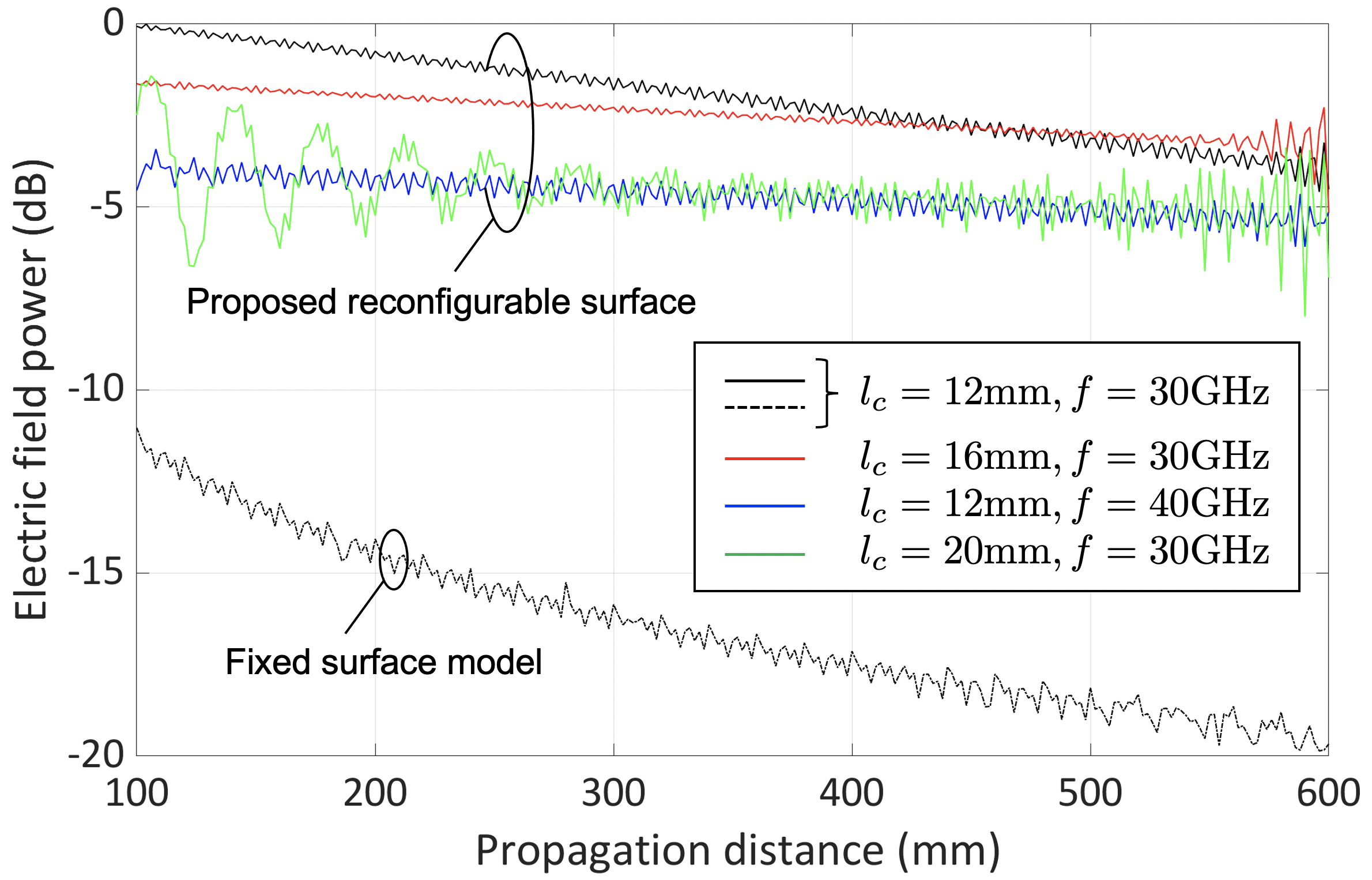}
\caption{The E-field in dB over propagation distance.}\label{fig:PLresults}
\end{figure}

\section{Simulation Results}
Simulations were conducted by CST Studio Suite 2020 and Taconic TLY-5 ($\varepsilon_r=2.2$, $\tan\delta = 0.0009@10{\rm GHz}$) was chosen as the dielectric layer and copper (annealed) as the metal layer. Galinstan with conductivity $\sigma_g$ was employed as the liquid metal alloy to fill the cavities when needed. Results in Fig.~\ref{fig:layer} demonstrate the E-field distribution over the surface when a direct path is formed by layers of cavities filled with Galinstan. We can see that the Galinstan layers are able to concentrate the field inside the path and have little leakage outside the path. In particular, there is some $25{\rm dB}$ difference of the field strength between inside and outside the path. Additionally, adding more layers of Galinstan filled cavities can result in better signal localization, suggesting that the proposed reconfigurable surface could be instrumental to the realization of controllable surface waves.

Fig.~\ref{fig:PLresults} provides the results of the power over propagation distance. First, we see that the power loss for the proposed surface is much less than the surface with no cavities filled with Galinstan. At a distance of $600{\rm mm}$, a $15{\rm dB}$ gain can be obtained using the proposed surface. Results also show that different channel widths and frequencies lead to different performance but the Galinstan filled cavities appear to keep the loss mild. Note that the curves look bumpy as numerous reflections occur when the waves interact with the cavities.


\section{Conclusion}
This paper proposed a reconfigurable surface platform in which cavities can be filled to form controllable surface wave pathways. The results have demonstrated great promises for enhancement and localization of surface waves.




\end{document}